# 3D Holoscopic Imaging for Cultural Heritage Digitalisation

Taha Alfaqheri *Taha.Alfaqheri@brunel.ac.uk*
Seif Allah El Mesloul Nasri *seifallah2020@gmail.com*
Abdul Hamid Sadka *Abdul.Sadka@brunel.ac.uk*

*Abstract:-*The growing interest in archaeology has enabled the discovery of an immense number of cultural heritage assets and historical sites. Hence, preservation of CH through digitalisation is becoming a primordial requirement for many countries as a part of national cultural programs. However, CH digitalisation is still posing serious challenges such as cost and time-consumption. In this manuscript, 3D holoscopic (H3D) technology is applied to capture small sized CH assets. The H3D camera utilises micro lens array within a single aperture lens and typical 2D sensor to acquire 3D information. This technology allows 3D autostereoscopic visualisation with full motion parallax if convenient Microlens Array (MLA) is used on the display side. Experimental works have shown easiness and simplicity of H3D acquisition compared to existing technologies. In fact, H3D capture process took an equal time of shooting a standard 2D image. These advantages qualify H3D technology to be cost effective and time-saving technology for cultural heritage 3D digitisation.

______________________________________________________*****______________________________________________________

## 1. Introduction:

Cultural Heritage is a key factor for maintaining the communities' identity and ensuring cultural and cognitive communication between generations, hence the importance of preserving the diverse types of CH. In fact, Museums have appeared as an effective mean for preserving the CH. However, they can just offer a limited service compared to the massive number of the CH assets that could be exhibited. This is coupled with the potentially high risks of loss and damage in case of accident or natural disasters, e.g. the collapse of 30 shelf-kilometers of the Cologne historical archive in 2009 due to maintenance work on an underground line and the fire at Herzogin Anna Amalia library in 2004. CH digitisation is the best solution to overcome these limitations. Moreover, it offers a larger and easier access to users from different sectors. In fact, the documentation of cultural assets is inherently a multimedia process, addressed through the digital representation of shape, appearance and preservation conditions of the Cultural Heritage (CH) object. CH assets are not cloneable physically or impeccably restorable, and hence their curation, as well as long-term preservation, require leveraging of advanced 3D modelling technologies. However, this could pose serious challenges since generating high-quality 3D models is still very time-consuming and expensive. Furthermore, the modelling is carried out for individual objects rather than for entire collections. For instance, the acquisition time of middle-sized objects ($50 \times 50 \times 50$ cm) requires up to 20 hours with structured light acquisition techniques according to 3D-COFORM Project [1].

The acquisition time and cost rise depending on both size and complexity of the target asset. Therefore, a cost-effective acquisition technology is highly required. Recently, a Holoscopic 3D camera [2] has been developed simulating the fly's eye technique which uses coherent replication of light to construct a true 3D scene in space. The developed H3D camera is a single-aperture DSLR camera with a transformed optical composition, where the primary added element is the micro lens array which allows capturing light field from different perspectives. This technique offers an enriched playback representation of the captured scene or objects including stereo images, multiview images, images with different depth levels and a complete Holoscopic image representation.

In this paper, 3D Holoscopic camera has been applied for CH assets acquisition for small sized assets within a designed environment and test conditions. The rest of this article is organised as follows: Section 2 addresses the relevant 3D acquisition techniques applied in CH domain and summarises their principles, advantages and drawbacks. The H3D camera is described in Section 3. The experimental results of H3D camera acquisition sessions are presented and discussed in Section 4. Finally, some concluding remarks are given in section 5.

## 2. Overview of the 3D acquisition system applied in CH

The first stage in CH digitisation process is the 3D image acquisition stage. It can be defined as the generation of different types of 3D spatial-temporal models. In this section, a brief overview of 3D image acquisition methods is discussed. Moreover, the advantages and drawbacks of each technique are presented focusing mainly on optical acquisition as a most popular acquisition technology in the CH sector. The CH assets can be acquired using different types of sensory devices, regarding the targeted application [3]. The choice of sensors also depends on the best trade-off between the required 3D reconstruction quality, cost, and acquired object sizes [3]. The 3D image acquisition systems are classified into two main categories, namely contact and non-contact acquisition systems (Figure 1) [4].

A contact acquisition system uses a mechanical arm to touch and record surface points of CH assets body. It is manually controlled by a high skilled worker or by automotive machine. Although it gives accurate results of the 3D reconstructed model (up to micron units), it is a costly technique, and it takes lots of time to acquire a medium size CH asset. A good example of such system is the Coordinate Measuring Machine (CMM) [5]. Clive in [6], presents a detailed evaluative study on different contact acquisition methods and its challenges. Shen and Moon proposed method can be used to accurately identify the reference ball centre in the performance test of touch trigger probes [6].

A contact acquisition system can be categorised into active and passive systems depending on the used touch probe mechanism. The active contact acquisition system works on electrical linear drive probe, whereas, the passive system uses a mechanical spring to produce probe force on object surface [7].

In non-contact acquisition systems, the 3D image data is recorded from the object surface and uses either light, magnetic field, or sound sources in the acquisition process. Then the system applies different image analysis and range finding methods to measure the position and distance of acquired surface points [5].

There are two types of non-contact acquisition systems, namely active and passive systems (Figure 1). The non-contact active system requires a patterned light source for the acquisition process to be projected on an object surface, for example, a 3D scanner system. However, passive systems do not require a patterned energy source for data acquisition [5].

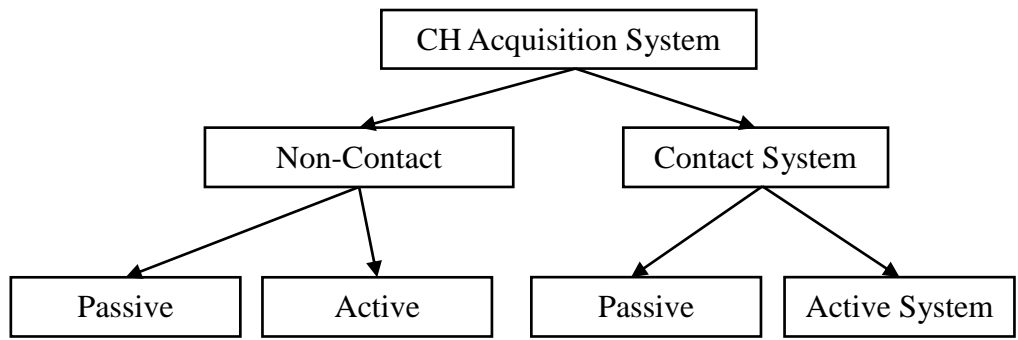

Figure 1. 3D image acquisition systems

The Optical acquisition techniques are a type of Non-contact passive system, and they offer a fast acquisition process compared to contact acquisition techniques, as no surface-touch with the assets is required. The optical acquisition techniques are classified into five different categories as shown in Figure 2 [4]. The efficiency of these techniques depends on the accuracy of surface points measurements and also on the used image analysis methods at post processing stage.

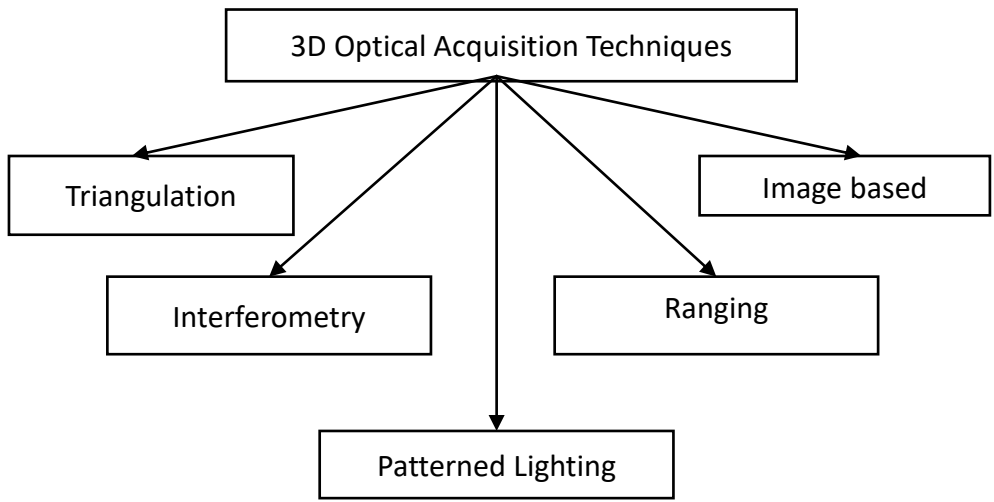

Figure 2. 3D Optical Acquisition Techniques

The triangulation technique projects a controlled light beam on the object surface and uses a sensor device to record the reflected light beam. The acquired 3D raw data is obtained by using disparity calculation between projected light and acquired image [7]. Although this technique acquires the object in short time, it is still suffering from limited accuracy and low resolution compared with other optical methods. Mojarrad implemented laser source in triangulation acquisition process for more precise object surface measurements [8].

Yang et al. introduced 3D reconstruction method using cloud points approach to give high accuracy surface points measurements with low complexity. However, this approach covers only large scale monitoring [9].

In the ranging technique, Laser and pulsed energy beam is used for distance measurements between the sensing device and object surface. The Interferometry technique depends on usingwavelength calculation which provides more accurate distance measurements compared to the ranging technique [8]. Patterned lighting technique includes projecting patterned light on an object surface; the reflected pattern is recorded and analysed for coordinates measurements of object surface points. It can capture complex details in one scan. However, this technique takes more calculations for extracting surface points positions compared with the previous two optical techniques (Triangulation and Interferometry).

Finally, Image-based method includes image analysis and features extraction of captured 2D image such as high-gradient points, shadows and object edges etc, then generates different 3D models [4].

The image-based method is a passive non-contact system as it does not require any patterned or structured light source in the acquisition process. It uses image analysis to obtain position coordinates of the object surface. The image-based technique reduces the acquisition complexity compared to the aforementioned methods as it requires only one shot to scan complex object surface [5].

With this technique, the real 3D object can be reconstructed and visualised with high details and varieties of different complex materials. Moreover, it acquires different objects sizes starting from small objects (9-18) cm to large objects such as historical sites. Snavely et al. in [11] presented a novel 3D interface with interactive browsing system for the unstructured scene. Another example, Sinha et al. proposed automatic 3D structure implementation using 2D interaction with multi view geometriccharacteristics [12].

With the advancement in the manufacturing of image camera sensors, image-based for close range acquisition is becoming flexible and adaptive to a wide variety of object sizes and types, from small objects to large buildings and archaeological sites. Moreover, it improves the acquisition processing time and becomes a more cost effective technique [10].

Nowadays, more research focuses on applying the image-based acquisition technique to satisfy the CH 3D digitalization requirements [11] which are: Visual 3D model quality, system portability and ease of use, acquisition cost and processing time.

### 3.Holoscopic 3D imaging, proposed technology for CH reservation

Holoscopic 3D imaging was initially proposed in 1908 by Lippmann[14]. Also known as integral or/and light field imaging, H3D imaging simulates the principle of Fly's eye technique by using a regularly spaced MLA within a single aperture camera. Each micro lens captures a slightly different perspective of the real scene (see figure 3a), offering an enhanced depth feeling and rich motion parallax without wearing any specific glasses.

The Holoscopic 3D imaging system is capable of creating and replicating a volumetric optical model of the recorded object in the form of a colourful dense light [15]. This coherent light is then constructed as a true 3D scene in space as shown in Figure 3b.

Figure 3 shows the fundamental principle of the H3D system. Light field of the real object is recorded as a 2D planar image through the MLA sheet. For the playback side, a convenient MLA is placed in front of the display screen allowing a light field diffusion to reconstruct the object in space [16].

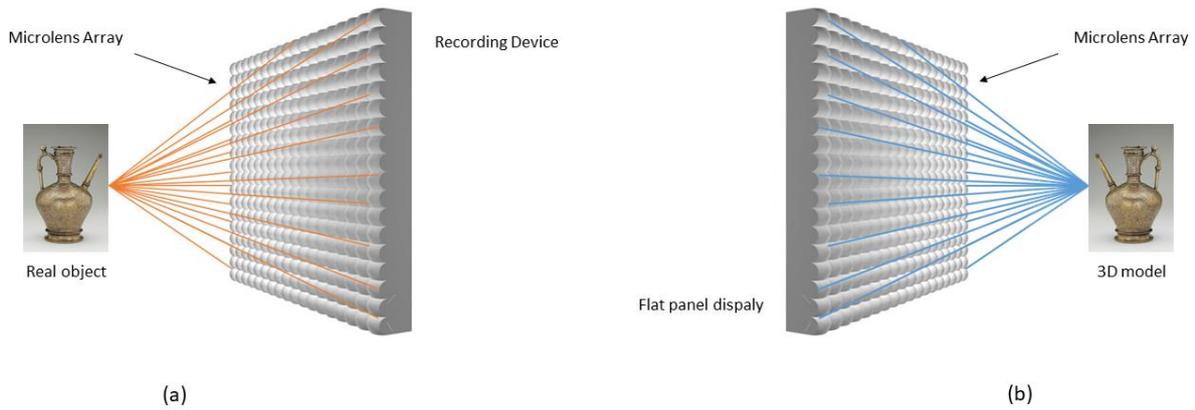

Figure 3. 3D holoscopic imaging scheme, (a) Capture, (b) Replay

Although the system in figure 3 presents a simple method for 3D reconstruction in space, it does not possess any depth control. The constructed image is inverted in depth (pseudoscopic). Also, objects far from the MLA will suffer from poor spatial representation. To overcome these limitations, an optical setup has been proposed in [15].

Figure 4 describes the optical components of the proposed camera. Compared to the setup in figure 3, the camera in figure 4 has two additional elements which are the objective and the relay lens. Further details about the camera architecture areprovided in the next section.

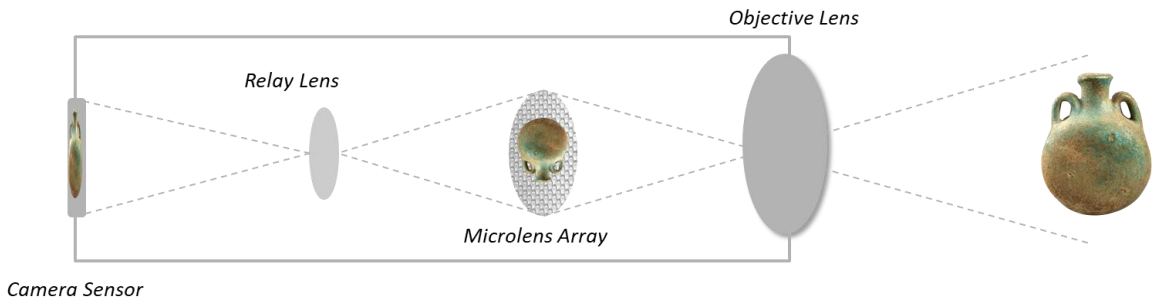

Figure 4. 3D Holoscopic camera architecture.

Moreover, H3D images can be converted to three different types of image during the post processing stage which are: 2D images with refocused plans, stereo images for 3D stereoscopic display and Multiview images for auto-stereoscopic 3D display [17]. Ease of use and mobility in addition to the multiple enriched 3D formats qualify the H3D imaging system to compete with the existing techniques of CH 3D digitalization.

### 4. Experimental Work

The H3D camera usesthe concept of single aperture camera (see Figure 5), in which MLA is slanted for few degrees to remove the moiré effect at H3D display and increase the horizontal spatial resolution at display output. The lenslet has a square shape with a size of 250×250 μm. The square aperture (SA) is used to reduce optical vignetting appearing in the resulting image corners. The prime lens offers better focus depth to the selected object in the scene with higher resolution colour fidelity. Extension rods are used to link the optical camera components and facilitate their replacement for further experiments and research.

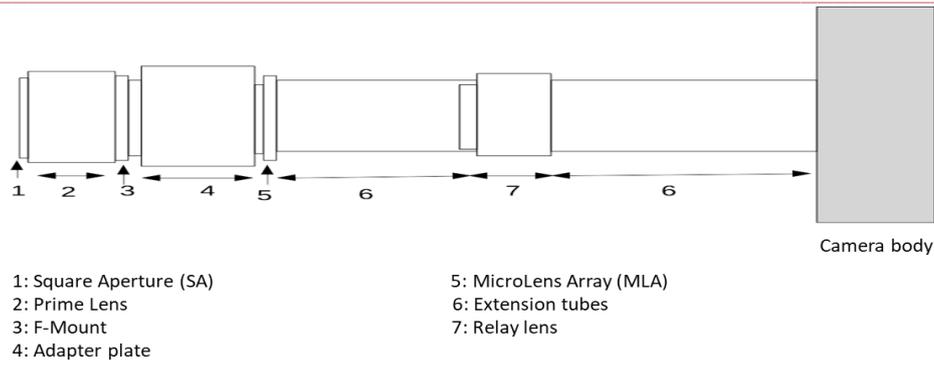

1: Square Aperture (SA)
2: Prime Lens
3: F-Mount
4: Adapter plate
5: MicroLens Array (MLA)
6: Extension tubes
7: Relay lens

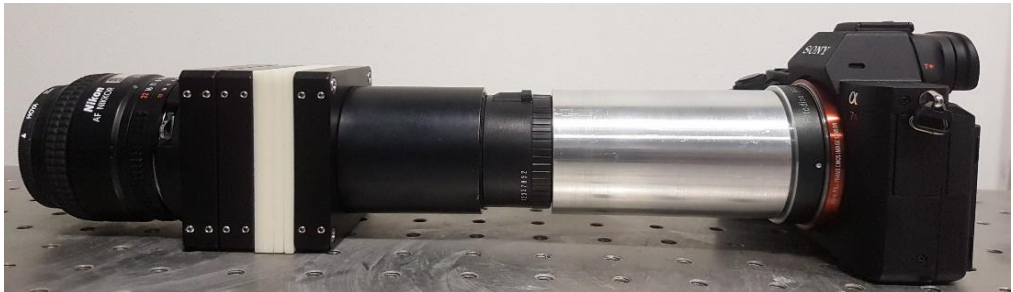

Figure 5. H3D Camera components

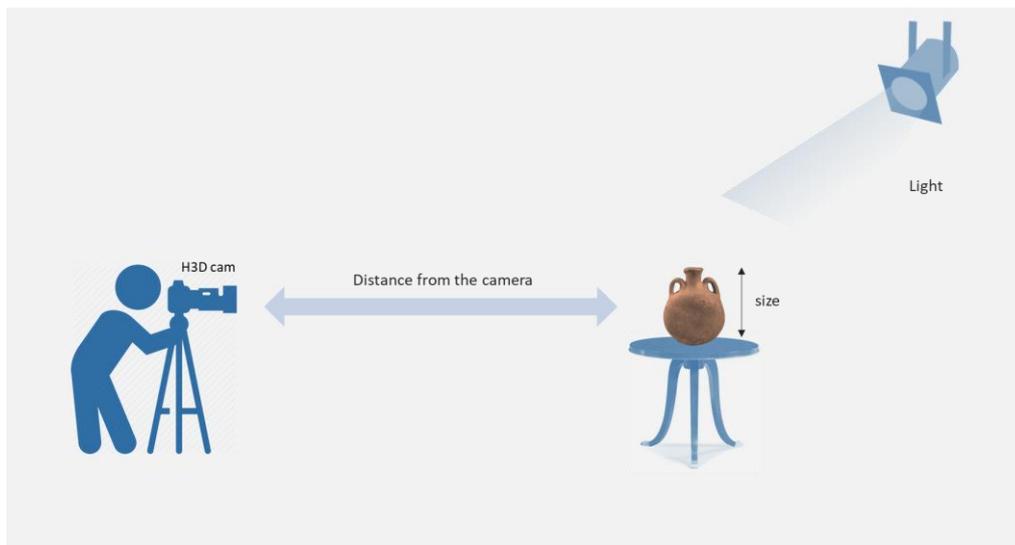

Figure 6. H3DI Capture process for small size cultural heritage assets

3DH images shooting environment and setup parameters are presented in this section. In addition, obtained raw holoscopic 3D data are presented and discussed. The experimental work was based on using a small sized asset, wherefore a red statuette of (11×5×3 cm) was utilised in these experimental tests, representing a small scale cultural heritage asset.

Figure 6 simulates the capture stage of CH assets. This stage is mainly depending on intrinsic H3D camera parameters and the surrounding external conditions. Before image acquisition, H3D camera parameters should be fixed and selected according to the shooting studio. The main intrinsic camera parameters are ISO, shutter speed and the focal length of both relay and prime lenses. These three parameters control the total amount of light entering the camera sensor.

In order to ensure that the full object appears clearly within the H3D frame, the distance between asset and camera should be selected carefully depending on the object size. Furthermore, a light projector isused to enhance the object brightness and ensure a balanced light distribution towards all parts and corners of the object.

Table 1 reports the intrinsicH3D camera parameters. The configuration has been set after a considerable number of repeated tests.These parameters offer a clear and bright view with focus points on the target object. The prime lens 50 mm is also referred to normal lens, it mimics the view field of the human eyes which will contribute later to having a better and natural perception of the holoscopic3D object. The lens' focal length is set to its largest value (1.4) allowing the acquisition of a maximum light field. Likewise, focal length of the relay lens is set to the highest value (2.8). Exposure time or shutter speed is the duration of time when the camera's shutter is open exposing the sensor to light. 1/30 seconds is chosen as the convenient exposure time for the shooting process. ISO value or film speed determines the camera sensor sensitivity to light.

Table 1 H3D camera intrinsic parameters

| Lens Aperture | 50 mm |
|---|---|
| Focal length (F) | 1.4 |
| Relay lens F | 2.8 |
| Exposure | 1/30 |
| ISO | 400 |

Three main criteria have been defined to evaluate an H3D raw image and decide if it is accepted for further extraction processing or not: Object position within the frame, well distinguishable details, and good microimages replication.

The object should fully appear in the H3D frame with a size allowing to distinguish details such as small text calligraphy or line shapes. This will depend on the asset size and its distance from the camera in addition to object surface characteristics. The microimage in the raw H3D image should replicate small parts of object scene with slightly different view angles. 3D information can be generated later from the replicated microimages. To achieve the desired replication, a suitable camera calibration should be done by optimising the combination between the light rays and the H3D camera optical components. Figure 7 presents an example where a wrong calibration has been applied. Although the object fully appears within the frame, the replicated microimages are blurred, and its details are not clear (Figure 7. b).

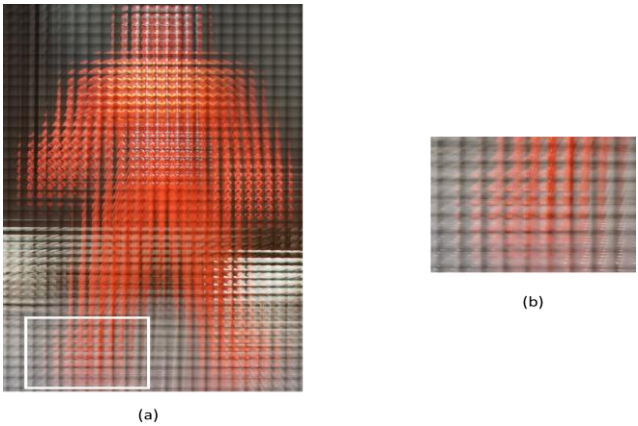

Figure 7. Raw H3D Image captured from 70 cm: (a) Full body sample (b) Magnified blurred foot

Table 2 summarises the conducted experimental tests regarding light intensity and asset distance from the H3D camera. The distance depends on the asset size. 70 cm has been selected as the most suitable length between the statuette and the H3D camera prime lens. Before this distance, the asset does not fully appear on the H3D image plan. Beyond 70 cm, the object will be far, and its details are not clearly perceived as shown in Figure 8.

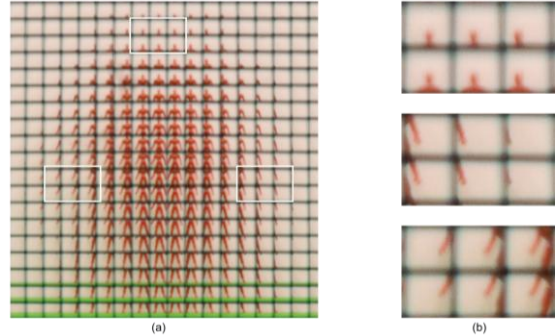

Figure 8. Raw H3D Image captured from 95 cm: (a) Full body sample (b) Magnified different parts of the image in part (a)

The light intensity around the asset is measured by a light metre in lumens/m2. It is controlled by different power levels of the light projector.

Finest values are reported regarding the different distances in table 2. Test number 5 represents the best combination of the light intensity and asset distance from the camera. Tests from one to four give good results in terms of visual quality of microimages. However, the acquired asset does not fit within the frame. The H3D resulted image is shown in figure 9.

Table 2 Environment configuration for H3D acquisition

| Test Number | Distance | Light Intensity (lumens/m2) |
|---|---|---|
| 1 | 30 | 117 |
| 2 | 40 | 126 |
| 3 | 50 | 131 |
| 4 | 60 | 133 |
| 5 | 70 | 150 |

Figure 9 shows H3D raw image captured using the H3D camera. A red statuette used as a sample of small object size, figure 9-a shows a full statuette using configuration parameters of test number (5) in table 2. Figure 9-b shows magnified sections of the head, right hand, and left hand of a statuette, respectively. The recorded planar intensity distribution of the object can be replayed using a particular lenticular sheet placed in front of the display screen. The design of lenticular sheet should be compatible with the implemented MLA at the acquisition stage.

The resulted raw H3D image is composed of many microimages. The number of microimages depends on MLA

used in the H3D camera. As shown in Figure 9-b, the microimages replication can be clearly seen. However, these replicated microimages do not include the same data. Each microimage represents a slightly different perspective part of the scene.

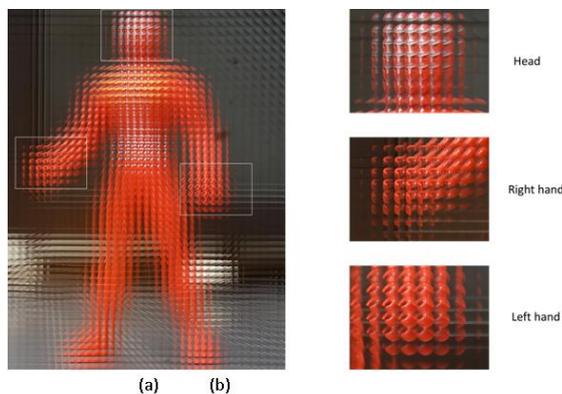

Figure 9. Raw H3D Image: (a) Full body sample (b) Magnified different parts of the image in part (a) showing replicated micro images with slightly different view angles

## 5. Conclusion:

This paper presented a novel H3D technology applied to acquire Cultural Heritage assets of small size. The study focused only on the acquisition stage of CH digitalization. The employed technology captures 3D information using only one single aperture camera. The H3D camera mimics the fly's eye concept by employing MLA, which records the CH assets from slightly different view angles.

The experimental work has been done with specific intrinsic camera settings and optical lens adaptation, in addition to different shooting distance and lighting conditions. The finding results proved that H3D is a time-saving technology as it just takes the time of an ordinary 2D image shooting. The presented raw H3D data can be extracted into multiple image formats (2D image, stereoscopic and Multiview) at a further stage. This work will push forward research on implementing and applying the light field technique within the CH sector. The future work will consider acquiring larger sized objects with adapted optical components.

## Acknowledgement:

This publication was made possible by NPRP grant 9-181-1-036 from the Qatar National Research Fund (a member of Qatar Foundation). The statements made herein are solely the responsibility of the authors.